\documentclass[pra,10pt,aps,twocolumn,amsmath,amssymb,altaffilletter]{revtex4-1}

\usepackage     [utf8]                  {inputenc}
\usepackage     [T1]                    {fontenc}
\usepackage                             {color}
\usepackage                             {graphicx}
\usepackage     [english]               {babel}
\usepackage     [colorlinks,citecolor=black,linkcolor=black,urlcolor=blue,bookmarks=false,hypertexnames=true]                        {hyperref}
\usepackage	[usenames,dvipsnames]	{pstricks}
\usepackage				{dcolumn}
\usepackage				{bm}
\usepackage				{textcomp}

\begin{document}

\title{Quantifying Stability of Quantum Statistical Ensembles}

\author{Walter Hahn}
\email{w.hahn@skoltech.ru}
\affiliation{Skolkovo Institute of Science and Technology, Skolkovo Innovation Centre, Nobel Street 3, Moscow 143026, Russia}
\affiliation{Institute for Theoretical Physics, Philosophenweg 19, 69120 Heidelberg, Germany}

\author{Boris V. Fine}
\email{b.fine@skoltech.ru}
\affiliation{Skolkovo Institute of Science and Technology, Skolkovo Innovation Centre, Nobel Street 3, Moscow 143026, Russia}
\affiliation{Institute for Theoretical Physics, Philosophenweg 19, 69120 Heidelberg, Germany}


\begin{abstract}
We investigate different measures of stability of quantum statistical ensembles with respect to local measurements. We call a quantum statistical ensemble ``stable'' if a small number of local measurements cannot significantly modify the total-energy distribution representing the ensemble. First, we numerically calculate the evolution of the stability measure introduced in our previous work \mbox{[Phys. Rev. E \textbf{94}, 062106 (2016)]} for an ensemble representing a mixture of two canonical ensembles with very different temperatures in a periodic chain of interacting \mbox{spins-\textonehalf}. Second, we propose other possible stability measures and discuss their advantages and disadvantages. We also show that, for small system sizes available to numerical simulations of local measurements, finite-size effects are rather pronounced.
\end{abstract}

\maketitle

\section{Introduction}
A long-standing problem for the foundations of quantum statistical physics is the proper choice of the initial energy distribution for the equilibrium description of isolated quantum systems. In particular, the use of narrow energy distributions, such as the conventional microcanonical distribution, does not have a clear justification in the quantum case~\cite{stab,pre_boris,fresch,fresch_i,fresch_ii,hantschel,wooters,brody_hughston,aarts_wetterich,eisert}. In our previous work~\cite{stab}, we proposed a solution to the above problem by introducing a criterion that physically realizable statistical ensembles describing a quantum system in equilibrium must be stable with respect to local measurements. We called a quantum statistical ensemble ``stable'' if a small number of local measurements cannot significantly modify the probability distribution of the total energy of the system~\cite{shimizu}. In order to introduce a quantitative stability criterion, we defined in Ref.~\cite{stab} a stability measure describing the modifications of the energy distribution caused by the measurements.

In this article, we numerically investigate the evolution of the above stability measure for a periodic chain of interacting spins-\textonehalf. We further show that the definition of the stability measure is not unique. We propose alternative stability measures and discuss them. We also explicitly show that, for small system sizes available to numerical simulations of local measurements, finite-size effects are rather pronounced. The numerical results obtained in this article are in good agreement with analytical estimates described in Ref.~\cite{stab}.

The rest of this article is organized as follows. In Sec.~\ref{sec_prelim}, we present a formulation of the problem, define the stability criterion and describe some results obtained in Ref.~\cite{stab}. In Sec.~\ref{sec_finite}, we investigate the finite-size effects of local measurements. In Sec.~\ref{sec_numres}, which is the main section of this article, we introduce alternative stability measures and compare them to the one introduced in Ref.~\cite{stab}.

\section{Formulation of the Problem and Preliminary Considerations} \label{sec_prelim}

\subsection{Hamiltonian}
The system of interest is a periodic chain of interacting spins-\textonehalf\ with nearest-neighbor interaction
\begin{equation} \label{eqn_ham2}
 {\cal H}=\sum_{i}J_x{S}_{ix}{S}_{(i+1)x}+J_y{S}_{iy}{S}_{(i+1)y}+J_z{S}_{iz}{S}_{(i+1)z},
\end{equation}
where ${S}_{ix}$, ${S}_{iy}$, and ${S}_{iz}$ are the spin projection operators for the lattice site $i$ on the axis $x$, $y$, and $z$, respectively, and \mbox{$J_x=-0.47$}, \mbox{$J_y=0.37$}, and $J_z=0.79$ are the coupling constants. We use $\hbar=1$. The characteristic single-spin energy for this Hamiltonian can be defined as $\epsilon_1\equiv(E_{\max}-E_{\min})/N_s$, where $E_{\min}$ is the minimal (ground-state) energy and $E_{\max}$ is the maximal energy of the system.

We investigate the above system by means of numerical simulations. For these simulations, we use the techniques described in Refs.~\cite{stab,my_thesis,tarek_prl,raedt_fast,dobrov}.

\subsection{Energy Distribution} \label{sec_endis}
The probability distribution $g(E)$ of the total energy $E$ is defined as $g(E)\equiv p(E)\nu(E)$, where $p(E)$ is the probability of occupying individual energy eigenstates with eigenenergy $E$, and $\nu(E)$ is the density of states as a function of the total energy $E$. We define the average total energy of the system as
\begin{equation}
 E_{\text{av}}\equiv\int_{E_{\min}}^{E_{\max}} E\,g(E)dE,
\end{equation}
and the width of the energy distribution as
\begin{equation}
 w_{g}\equiv\sqrt{\int_{E_{\min}}^{E_{\max}} (E-E_{\text{av}})^2 g(E)dE}.
\end{equation}

\subsection{Local Measurement}
We consider random instantaneous projective measurements of individual spins of the system. In general, there are two effects of such measurements on the total-energy distribution $g(E)$~\cite{stab,my_thesis}: narrowing and broadening of $g(E)$. Narrowing originates from correlations between the total energy $E$ and the measurement outcomes. Due to these correlations, the outcome of a local measurement typically yields partial information about the total energy $E$ and, therefore, the post-measurement $g(E)$ is narrower than the pre-measurement $g(E)$. Broadening of $g(E)$ is caused by the off-diagonal elements of the projection operators describing the measurements in the basis of the total-energy eigenstates. This broadening effect also leads to heating, which we define as the drift of the average energy $E_{\text{av}}$~\cite{stab,my_thesis}. For macroscopic system sizes and broad $g(E)$, the heating and the broadening effects are estimated to be negligible in comparison with the narrowing effect~\cite{stab}. However, for finite systems, the heating and the broadening are significant and complicate the analysis of simulation results, see Sec.~\ref{sec_finite}.

We limit our investigations in this article to ideal projective measurements because they are easier to treat numerically. For more general classes of measurements, though, we expect qualitatively similar results.

\subsection{Stability Criterion}
Following Ref.~\cite{stab}, we introduce the following stability criterion: {\it A physically realizable quantum statistical ensemble describing a stationary state of a macroscopic system must be stable with respect to a small number of any arbitrarily chosen local measurements within the system. The number of measurements $n$ is called small, if $n \ll \sqrt{N_s}$. The ensemble is called stable, if}
\begin{equation} \label{eqn_dg}
 \Delta G(n) \equiv \int_{E_{\min}}^{E_{\max}} \left| g_n(E) - g_0(E) \right| dE\ll1,
\end{equation}
where $g_0(E)$ is the initial energy distribution (before the measurements) and $g_n(E)$ is the energy distribution after $n$ measurements. The above stability measure $\Delta G(n)$ must be averaged over possible measurement outcomes. We denote the result of this averaging as $\overline{\Delta G}(n)$. In contrast to Ref.~\cite{stab}, we consider in this article small system sizes $17\leq N_s\leq 25$ instead of macroscopic systems and, therefore, the above requirement for the number of local measurements $n\ll\sqrt{N_s}$ must be relaxed. Here, we consider $n\leq6$.

The definition of the stability measure $\Delta G(n)$ given in Eq.~\eqref{eqn_dg} is not unique. In Sec.~\ref{sec_numres}, we propose other possible stability measures and compare them to the above $\Delta G(n)$.

\section{Broadening and Heating Effects for Finite Systems} \label{sec_finite}

\begin{figure}[t]
	\centering
	\includegraphics[width=0.48\textwidth]{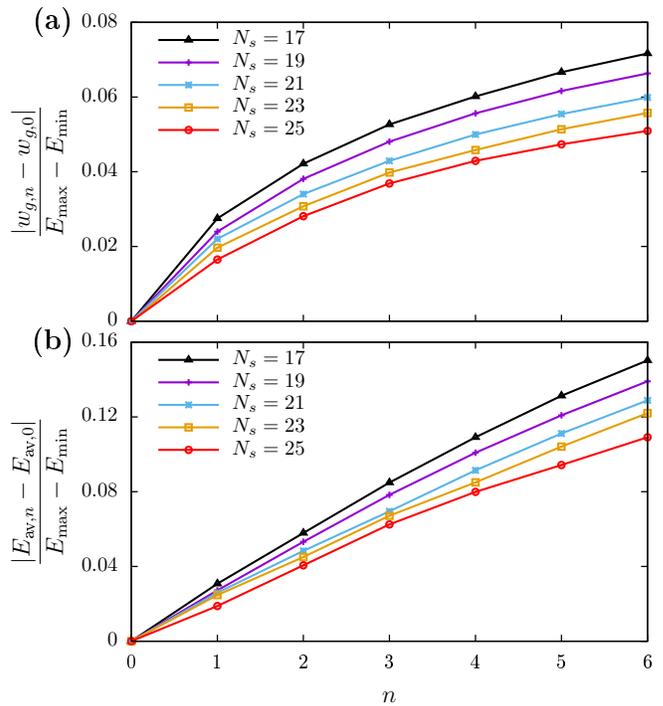}
	\caption{Numerical simulations of the broadening and heating effects for a periodic chain of $N_s$ spins-\textonehalf\ with initial canonical energy distribution $g_0(E)\cong\exp(-E/T_1)\nu(E)$, where $T_1=0.1$. (a) Broadening effect quantified by the relative change of the width $|w_{g,n}-w_{g,0}|/(E_{\max}-E_{\min})$ and (b) heating effect quantified by the relative change of the average energy $|E_{\text{av},n}-E_{\text{av},0}|/(E_{\max}-E_{\min})$ as a function of the number of local measurements $n$. Symbols represent numerical calculations, lines are guides to the eye.}
	\label{fig_finite_size}
\end{figure}

Let us now investigate the broadening and the heating effects of local measurements. We study these effects for the initial Gibbs energy distribution $g_0(E)\cong\exp(-E/T_1)\nu(E)$, where $T_1=0.1$ and $k_\text{B}=1$. For each measurement, we choose the lattice site and the measurement axis randomly. The time delay between two successive measurements is chosen randomly from interval $[0,2]$.

We characterize the broadening effect by the relative change of the width $|w_{g,n}-w_{g,0}|/(E_{\max}-E_{\min})$, and the heating effect by the relative deviation of the average energy from its initial value $|E_{\text{av},n}-E_{\text{av},0}|/(E_{\max}-E_{\min})$. The results shown respectively in Fig.~\ref{fig_finite_size} (a) and (b) indicate significant heating and broadening, which are both the finite-size effects. After $n=6$ local measurements, $|w_{g,n}-w_{g,0}|\sim E_{\max}-E_{\min}$ and $|E_{\text{av},n}-E_{\text{av},0}|\sim E_{\max}-E_{\min}$, which corresponds to a significant modification of $g(E)$. At the same time, the plots in Figs.~\ref{fig_finite_size} (a) and (b) also indicate that the above finite-size effects become weaker for larger system sizes $N_s$ in agreement with the analytical estimates of Refs.~\cite{stab,my_thesis}.

\section{Investigation of Stability Measures} \label{sec_numres}

\begin{figure}[b]
	\centering
	\includegraphics[width=0.48\textwidth]{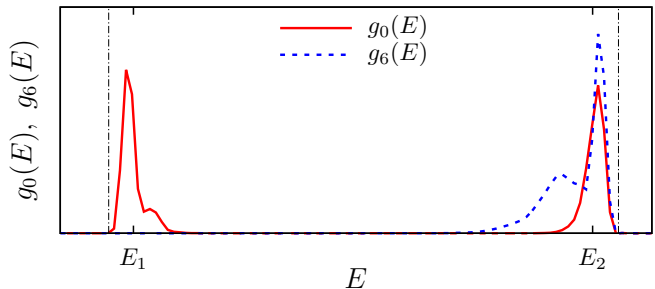}
	\caption{Initial energy distribution $g_0(E)$ defined by Eq.~\eqref{eqn_init} (solid red line) and the typical energy distribution $g_6(E)$ after 6 single-spin measurements or, equivalently, 3 NN-measurements (blue dashed line). The dash-dotted vertical lines indicate $E_{\min}$ and $E_{\max}$.}
	\label{fig_example}
\end{figure}

In the following, we consider several possible measures of stability with respect to local measurements, and compare their performance in a setting when the interacting spin chain is initially characterized by a mixture of two canonical ensembles of very different temperatures $T_1$ and $T_2$, where $T_1=-T_2=0.1$ (The concept of negative temperature is routinely used in systems with limited energy range per particle. It is applicable in particular to both classical and quantum spin systems~\cite{purcell1951,negative_temp,astrid,dunkel,frenkel,stab}). The corresponding energy distribution is
\begin{equation} \label{eqn_init}
 g_0(E)=\frac{1}{2}\left(A_1\exp\!\left[-\frac{E}{T_1}\right]+A_2\exp\!\left[-\frac{E}{T_2}\right] \right)\nu(E),
\end{equation}
where $A_1$ and $A_2$ are normalization constants determined from the conditions $\int A_1\exp\left[-E/T_1\right]\nu(E)dE=1$ and $\int A_2\exp\left[-E/T_2\right]\nu(E)dE=1$. 

The initial energy distribution $g_0(E)$ has two narrow peaks around the average energies $E_1$ and $E_2$ for the contributing canonical ensembles with temperatures $T_1$ and $T_2$, respectively - see Fig.~\ref{fig_example}. Since \mbox{$E_2-E_1\sim E_{\max}-E_{\min}$}, we call the above $g_0(E)$ broad. Such a distribution belongs to a broader class of fragile quantum states defined in Ref.~\cite{fragile}. The narrowing effect for the two-peak distribution $g_0(E)$ manifests itself in the the suppression of one peak: either the peak at $E_1$ or the peak at $E_2$, dependent on the measurement outcomes.

For the interacting spin system~\eqref{eqn_ham2}, at least two accidental measurements sufficiently close in space are required to induce the narrowing of $g(E)$~\cite{stab,my_thesis}. Therefore, in order to shorten the simulation times necessary to observe the ensemble-narrowing effect, we implement random nearest-neighbor (NN-) measurements. This means that the spin site for the odd-numbered measurements is chosen randomly, while, for even-numbered measurements, a nearest neighbor of the previously measured spin is chosen. The time delays between individual single-spin measurements are selected randomly from interval $[0,2]$. To make the narrowing effect even more pronounced, the spins are measured along the direction of the strongest interaction ($z$ axis in our case), for which the correlation between the outcome of a NN-measurement and the total energy $E$ is expected to be the strongest. Since $|J_z|>|J_x|,|J_y|$, and $J_z>0$, the dominant spin configuration at temperature $T_1$ (peak at $E_1$) is anti-ferromagnetic, whereas, at temperature $T_2$ (peak at $E_2$), it is ferromagnetic.

\subsection{Stability Measure given by Eq.~\eqref{eqn_dg}}

\begin{figure*}[t]
\includegraphics[width=\textwidth]{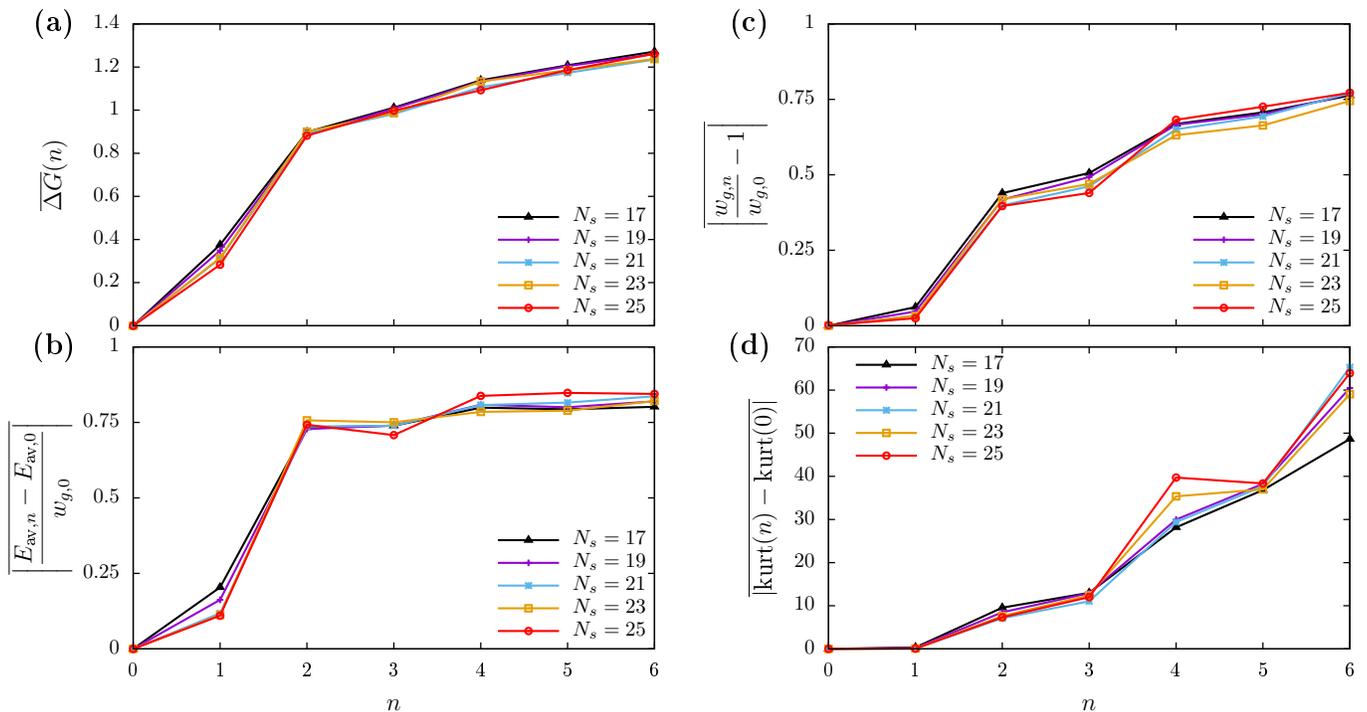}
\caption{Stability measures as functions of the number $n$ of local measurements for the initial energy distribution $g_0(E)$ in Eq.~\eqref{eqn_init} for a periodic chain of $N_s$ spins-\textonehalf\ with Hamiltonian~\eqref{eqn_ham2}: (a) $\overline{\Delta G}(n)$ defined in Eq.~\eqref{eqn_dg}, (b) relative change of the average energy $\left|E_{\text{av},n}-E_{\text{av},0}\right|/w_{g,0}$, (c) relative change of the width $|w_{g,n}/w_{g,0}-1|$, (d) change of the kurtosis $|\text{kurt}(n)-\text{kurt}(0)|$. Symbols represent numerical calculations, lines are guides to the eye.}
\label{fig_other}
\end{figure*}

Let us first investigate the stability measure $\Delta G(n)$ defined by Eq.~\eqref{eqn_dg}. The numerically computed behavior of $\overline{\Delta G}(n)$, where the bar denotes the average over measurement outcomes, is plotted in Fig.~\ref{fig_other} (a). This plot indicates that, after a few measurements, $\overline{\Delta G}(n)\sim1$ independent of $N_s$, which implies that the two-peak energy distribution $g_0(E)$ in Eq.~\eqref{eqn_init} is unstable with respect to local measurements, which in turn is consistent with the analytical estimates of Refs.~\cite{stab,my_thesis}. For $n\leq3$, the change of $\overline{\Delta G}(n)$ at odd-numbered measurements is smaller than at even-numbered measurements. The reason is that each even-numbered measurement completes a NN-measurement.

According to Fig.~\ref{fig_other} (a), $\overline{\Delta G}(n)$ grows above 1 which is, at first sight, surprising. Typically, local measurements lead to the suppression of one of the two peaks as shown in Fig.~\ref{fig_example}. If this were the only effect, $\overline{\Delta G}(n)$ would be at most 1. The above discrepancy is caused by the broadening and the heating, cf. Fig.~\ref{fig_example}.

The results obtained imply that $\overline{\Delta G}(n)$ is prone to finite-size effects. An advantage of $\overline{\Delta G}(n)$ is that it accounts for all possible modifications of the energy distribution $g(E)$. We also note here that the values of $\overline{\Delta G}(n)$ depend on the partition of the energy axis into bins~\footnote{In our calculations, we divided the energy interval $[E_{\min},E_{\max}]$ into bins of equal width. The number of these bins was of the order of 100.}.

\subsection{Alternative Stability Measures}
In the following, we define alternative stability measures and discuss them.

One such an alternative stability measure is the relative deviation of the average energy from its initial value $\left|E_{\text{av},n}-E_{\text{av},0}\right|/w_{g,0}$. Initially, $E_{\text{av},0}$ is centered between the two peaks. If one peak becomes suppressed due to the measurements, $E_{\text{av},n}$ jumps to $E_{\text{av},n}\approx E_1$ or to $E_{\text{av},n}\approx E_2$, which in both cases imply $\left|E_{\text{av},n}-E_{\text{av},0}\right|/w_{g,0}\approx1$.
 
A possible stability criterion is
\begin{equation} \label{eqn_alt1}
 \overline{\left|\frac{E_{\text{av},n}-E_{\text{av},0}}{w_{g,0}}\right|}\ll1.
\end{equation}

The calculated values of $\left|E_{\text{av},n}-E_{\text{av},0}\right|/w_{g,0}$, shown in Fig.~\ref{fig_other} (b), indicate that this stability measure is similar to $\Delta G(n)$. A disadvantage of $\left|E_{\text{av},n}-E_{\text{av},0}\right|/w_{g,0}$ as a stability measure is that it only accounts for the average energy $E_{\text{av},n}$ and, therefore, does not include modifications of $g(E)$ which do not change $E_{\text{av},n}$.
 
Another alternative stability measure is the change of the relative width $\left|w_{g,n}/w_{g,0}-1\right|$. Initially, $w_{g,0}\approx E_2-E_1\sim \epsilon_1 N_s$. In the case of a complete suppression of one peak, the final width is $w_{g,n}\ll\epsilon_1 N_s$.
 
The stability criterion for this measure is
\begin{equation}\label{eqn_alt2}
 \overline{\left|\frac{w_{g,n}}{w_{g,0}}-1\right|}\ll1.
\end{equation}

The calculated values of $|w_{g,n}/w_{g,0}-1|$ are shown in Fig.~\ref{fig_other} (c). The overall behavior of $|w_{g,n}/w_{g,0}-1|$ indicates that this stability measure is similar to $\Delta G(n)$. A possible disadvantage of $|w_{g,n}/w_{g,0}-1|$ as a stability measure is that it is based on the width $w_{g,n}$ only and does not account for modifications of $g(E)$ which do not change the width.

A further alternative stability measure is the kurtosis $\text{kurt}(n)\equiv M_{4,n}/M^2_{2,n}$, where $M_{2,n}$ and $M_{4,n}$ are the second and the fourth moments of the distribution $g_n(E)$: $M_{2,n}\equiv w^2_{g,n}$ and $M_{4,n}\equiv\int (E-E_{\text{av},n})^4\ g_n(E)dE$. The kurtosis is a measure of the shape of $g(E)$. For example, for the Gaussian shape, the kurtosis is equal to 3. For a two-peak $g(E)$, the kurtosis is close to 1.
 
A possible stability criterion is
\begin{equation}\label{eqn_alt3}
 \overline{|\text{kurt}(n)-\text{kurt}(0)|}\ll1.
\end{equation} 

As shown in Fig.~\ref{fig_other} (d), $|\text{kurt}(n)-\text{kurt}(0)|$ grows to large values of the order of 50. These large values reflect the fact that the kurtosis is very sensitive to small variations of the tails of $g(E)$. At the same time, the kurtosis is not sensitive to the variations of $g(E)$ that do not change the overall shape - for example, given an initial broad $g_0(E)$ having a Gaussian shape, which becomes narrower without changing the shape, $\text{kurt}(n)$ would remain constant. Therefore, the kurtosis is less suitable for practical calculations than the other three stability measures.

Other alternative definitions of the stability measure may include more elaborate distances between probability distributions, such as the Monge-Kantorovich metric~\cite{monge,kantorovich,villani} or the Wasserstein metric~\cite{wasserstein}. In fact, the former has been already applied to quantum states in Ref.~\cite{zyczkowski}. The above measures may have advantages over the ones considered in the present work but they are more difficult to calculate in practice.

\section{Conclusions} \label{sec_conclusion}
We investigated the stability measure $\Delta G(n)$ introduced in Ref.~\cite{stab} and other proposed stability measures for quantum statistical ensembles with respect to local measurements. We showed that the definition of the stability measure is not unique. However, the stability measure which is suitable for any $g(E)$ is $\Delta G(n)$. The numerical results discussed in this article are in good agreement with the analytical estimates described in our previous article~\cite{stab}.

\textit{Acknowledgements} - This work was supported by a grant from the Russian Science Foundation (Project No. 17-12-01587). W.H. is grateful for the support from \textit{Studienstiftung des deutschen Volkes} during the initial phase of this work.

\end{document}